# Strong short-range magnetic order in a frustrated FCC lattice and its possible role in the iron structural transformation

*A. N. Ignatenko[1], A.A. Katanin[1,2], and V.Yu.Irkhin[1]*

[1]*Institute of Metal Physics, Russian Academy of Science, 620041 Ekaterinburg, Russia*

[2]*Max-Planck-Institute für Festkörperforschung, 70569 Stuttgart, Germany*

We investigate magnetic properties of a frustrated Heisenberg antiferromagnet with a face-centered cubic (FCC) lattice and exchange interactions between the nearest- and next-nearest neighbours, $J_1$ and $J_2$. In a collinear phase with the wave vector $\mathbf{Q} = (\pi,\pi,\pi)$ the equations of the self-consistent spin-wave theory for the sublattice magnetization and the average short range order parameter are obtained and numerically solved. The dependence of the Neel temperature $T_N$ on the ratio $J_2/J_1$ is obtained. It is shown, that at strong enough frustration there is a wide temperature region above $T_N$ with the strong short range magnetic order. Application of this result to description of structural phase transition between $\alpha$ and $\gamma$-phase of Fe is considered.

PACS: 75.10. Jm, 75.30. Kz, 75.50. Bb

Short range magnetic order (SRMO) can be important for description of thermodynamic properties in a vicinity of magnetic phase transitions. It influences also the structural phase transitions, since the presence of pronounced short-range magnetic order (even in the absence of long-range one) in one of the phases will affect conditions of appearance of a new (another) phase with changing temperature, pressure or concentration of impurities. In this respect, the vicinity of the magnetic quantum phase transitions (MQPT) occurring at temperature $T = 0$ and certain relation between physical parameters (exchange integrals, pressure, concentration of impurities etc.) is especially interesting, since close to MQPT the order parameter and temperature of transition are small, and there can be a wide temperature region of the strong short range order influencing structural transformations.

The abovementioned arguments are expected to be applicable, in particular, to iron. It is known, that it exists in two crystal phases: body-centered, (BCC, $\alpha$-Fe) and face-centered cubic (FCC, $\gamma$-Fe). At low temperatures the BCC phase, which is ferromagnetically ordered at $T < 1045$ K is stable. At $T = T_s = 1185$ K there is a structural transition to the $\gamma$-phase which is stable in an interval of temperatures about 200K above $T_s$. With further increase of temperature iron undergoes the transition to a high-temperature BCC-phase. The structural phase transition from $\alpha$ to a $\gamma$-phase may be related to the presence of strong short range antiferromagnetic order in $\gamma$-phase. Contrary to the macroscopic three-dimensional samples where the $\gamma$-phase is observed only at high temperatures in the magnetically disordered state, in granular- and precipitate samples it is stable at low temperatures and orders antiferromagnetically at a Neel temperature $T_N \sim 60$ K [1]. Despite so low value of transition temperature (in comparison to the temperature of ferromagnetic phase transition in the BCC phase), a strong short range antiferromagnetic order can exist at $T \gg T_N$. Experimentally this order could be observed in an inelastic neutron scattering experiments.

Previous theoretical investigations of two- [2] and quasi-two-dimensional [3] systems have shown, that SRMO indeed can exist in the temperature region $T_N < T < J$ ($J$ is the exchange interaction within a layer). Especially wide area of existence SRMO is expected near a frustration point (that is a point of strongest competition of various magnetic phases) where $T_N \ll J$, as was discussed previously for the square and simple cubic lattices with an exchange between nearest- and next-nearest neighbors [4]. On a strongly frustrated square lattice a state without the long range magnetic order (a spin liquid) is possible [5]. A similar situation probably takes place in the FCC lattice, where the frustration is expected to be stronger, than in the simple cubic lattice. In this connection investigation of magnetic properties of three-dimensional systems with FCC lattice near to a frustration point where SRMO is expected to exist at temperatures much larger than $T_N$ is of theoretical interest.

Convenient method of investigation of SRMO is the self-consistent spin - wave theory (SSWT) [2-4], where the short range order parameter

$$\gamma_\delta = \sqrt{|\mathbf{S}_i \mathbf{S}_{i+\delta}|}/S$$

($S_i$ is the spin operator at a site $i$, $\delta$ is a radius-vector of the nearest or next-nearest neighbours) arise naturally after decoupling of the fourfold terms of Bose operators corresponding to the magnon interaction. This theory is based on a spin - wave picture of an excitation spectrum which was shown to be applicable in a wide interval of temperatures above Neel temperature by the experimental data on systems with low $T_N$ (e.g. on perovskites). Contrary to the usual spin-wave theory (SWT), SSWT is applicable also in the paramagnetic phase and takes into account both temperature, and quantum fluctuations of SRMO.

As a model for spin subsystem here we consider the quantum Heisenberg model on the FCC lattice

$$H = \frac{1}{2}\sum_{ij} J_{ij} \mathbf{S}_i \mathbf{S}_j \qquad (1)$$

with the interaction between nearest- and next-nearest neighbours, $J_1$ and $J_2$, respectively. In limit $S \to \infty$ this model reduces to a classical Heisenberg model. The corresponding classical energy of possible spin configurations depending on the ratio $r = J_2/J_1$ ($J_1 > 0$) is shown on Fig.1.

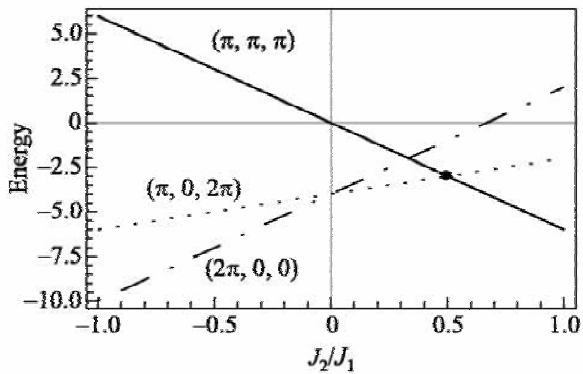

Fig. 1. Energy of various classical spin configurations depending on the ratio $r = J_2/J_1$

At $r > 1/2$ the collinear structure with wave vector $\mathbf{Q} = (\pi, \pi, \pi)$ is stable:

$$\langle \mathbf{S}_i \rangle = (-1)^{\mathbf{Q} \cdot \mathbf{R}_i} \mathbf{M} \qquad (2)$$

(the lattice constant $a=1$), where $\mathbf{M}$ is a vector of the sublattice magnetization, $\mathbf{R}_i$ is a radis-vector of site $i$ of a lattice. In the range $0 < r < 1/2$ the ground state becomes non-collinear with a wave vector $\mathbf{Q}=(\pi, 0, 2\pi)$ and

$$\langle \mathbf{S}_i \rangle = \mathbf{M}_1 \cos \mathbf{Q} \mathbf{R}_i + \mathbf{M}_2 \sin \mathbf{Q} \mathbf{R}_i \qquad (3)$$

where $\mathbf{M}_1$ and $\mathbf{M}_2$ are orthogonal vectors of unit length. At $r < 0$ the ground state again becomes collinear with a wave vector $(2\pi, 0, 0)$. In limit $S \to \infty$ all phase transitions between these phases are of the first order. With account of quantum fluctuations the point $r = 1/2$ (it is marked on the figure) appears to be a frustration point [6]. On the other hand, in a vicinity of a point $r = 0$, quantum fluctuations do not change qualitatively the ground state phase diagram and the paramagnetic phase does not arise.

To study a role quantum and temperature fluctuations near frustration point in collinear phase $\mathbf{Q} = (\pi,\pi,\pi)$, we use the Baryakhtar – Krivoruchko – Jablonskii representation of spin operators through Bose and Fermi operators $b_i$ and $c_i$ [7]:

$$S_i^z = (-1)^{\mathbf{Q} \cdot \mathbf{R}_i} : \left( S - B_i^\dagger B_i - (2S+1)c_i^\dagger c_i \right) :,$$
$$S_i^+ = \sqrt{2S} B_i, \qquad (4)$$
$$S_i^- = \sqrt{2S} : \left( B_i^\dagger - \frac{1}{2S} B_i^\dagger B_i^\dagger B_i \right) : - \frac{2(2S+1)}{\sqrt{2S}} B_i^\dagger c_i^\dagger c_i,$$

where $B_i = b_i$ on one sublattice, and $B_i = b^+_i$ on another. The symbol :: stands for normal ordering.

In a standard variant of the SSWT [2, 3] one applies the representation (4) and decouples the fourfold terms of the interaction within the Hartree-Fock approximation. For considering phase $\mathbf{Q} = (\pi,\pi,\pi)$ this variant meets however some difficulties, connected to the "order from disorder" phenomenon [8]. The reason is that in the limit $S \to \infty$ the ground state of the Hamiltonian (1) is degenerate at $r > 1/2$. It represents a system of four simple cubic sublattices with the chess-board antiferromagnetic order on each sublattice, and the direction of the sublattice magnetization in different sublattices is independent. In the standard SWT this circumstance results in that at $T < T_N$ new gapless excitations appear (in addition to standard Holdstone modes), which correspond to the rotation of Neel vectors of different sublattices with respect to each other. However if one considers corrections to SWT in $1/S$, in those points where the additional gapless excitations should exist, spurious "quantum" gaps open, which are growing with increase of temperature.

Such behaviour of gaps contradicts in particular the experiment on garnet $Ca_3Fe_2Ge_3O_{12}$ where similar situation takes place [9]. The erroneous growth of "quantum" gaps predicted by standard SSWT, is extremely important for the behavior of thermodynamic quantities as it leads to absence to strong overestimate $T_N$ in comparison with SWT and requires resummation of perturbation series.

This problem of standard variant of SSWT can be cured with introduction of the average short-range order parameter in which "quantum" gaps are absent at any temperatures. Average SRMO parameter is defined as follows [3]:

$$\bar{\gamma} = (\gamma_{1AF} - \gamma_{1F} + r\gamma_2)/r \quad (5)$$

where $\gamma_{AF}$ and $\gamma_F$ – respectively the SRMO parametres for the nearest antiferro - and ferromagnetically ordered neighbours, and $\gamma_2$ is the SRMO parameter for the next-nearest - neighbours. After substitution of (4) to the Hamiltonian and Hartree-Fock decoupling of the fourfold interaction we have the equations (here and hereafter we put $J_1 = 1$)

$$M^z = (S+1/2)\coth\left[\frac{e_0(S+1/2)}{T}\right] - \frac{1}{2}\int\frac{d^3k}{(2\pi)^3}\frac{P_k \coth(E_k/2T)}{E_k},$$

$$S\bar{\gamma} = (S+1/2)\coth\left[\frac{e_0(S+1/2)}{T}\right] + \quad (7)$$

$$+\frac{1}{2}\int\frac{d^3k}{(2\pi)^3}\left(g_k \cos\frac{k_x+k_y}{2} + rg_k \cos k_y - P_k \cos\frac{k_y+k_z}{2}\right)\frac{\coth(E_k/2T)}{E_k}$$

where $E_k = \sqrt{P_k^2 - g_k^2}$ is a spectrum elementary excitations,

$e_0 = 6Sr\bar{\gamma} - \mu,$

$$P_k = e_0 + 2S\bar{\gamma}\left(\cos\frac{k_y-k_x}{2} + \cos\frac{k_y+k_z}{2} + \cos\frac{k_x+k_z}{2}\right),$$

$$g_k = 2S\bar{\gamma}\left(\cos\frac{k_z-k_x}{2} + \cos\frac{k_z-k_y}{2} + \cos\frac{k_x+k_y}{2}\right) +$$

$$+2S\bar{\gamma}r(\cos k_x + \cos k_y + \cos k_z). \quad (8)$$

Here the Boson chemical potential $\mu$ is introduced; in the ordered phase $\mu = 0$, while in the paramagnetic phase $\mu<0$ is determined from the condition $\bar{S} = 0$ [2]. At the frustration point $r = 1/2$, in the assumption magnetically-ordered phase ($\mu = 0$), the spectrum of spin waves has the form

$$E_k = 2S\bar{\gamma}\left|\cos\frac{k_x}{2} + \cos\frac{k_y}{2} + \cos\frac{k_z}{2}\right| \times$$

$$\times\left|\sin\frac{k_x}{2} + \sin\frac{k_y}{2} + \sin\frac{k_z}{2}\right|. \quad (9)$$

This spectrum vanishes on a certain surfaces in the Brilloin zone (see Fig. 2), which results in that the magnetisation (6) at zero and final temperatures diverges to $-\infty$. This contradicts the assumption of magnetically ordered phase ($\bar{s} > 0$) and means, that as a result of competition of two magnetic phases $(\pi,\pi,\pi)$ and $(\pi,0,2\pi)$ a new phase with no long range magnetic order (a spin liquid) can occur near to a frustration point $r = 1/2$. Note, that, contrary to the FCC lattice, for simple cubic lattice a formal divergence of the sublattice magnetisation is absent near a frustration point at $T = 0$ (however it is present at $T>0$) as the spectrum of spin waves vanishes only on certain lines of the Brillouin zone in that case. Accordingly, the solution of SSWT equations with the average order parameter shows, that the appearance of spin-liquid on the SC lattice is possible only at $S = 1/2$ while in FCC to a lattice it appears at any $S$.

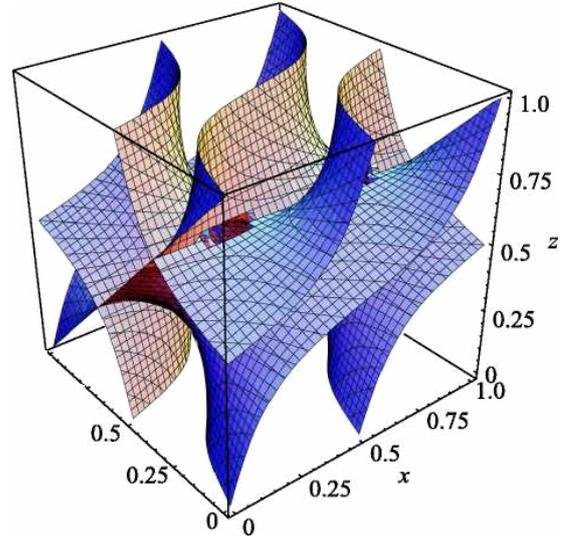

Fig. 2. Surfaces of zero of spin wave spectrum in Brillouin zone for $r = J_2/J_1 = 1/2$ in coordinates $x = (k_x+k_y)/(4\pi)$, $y = (k_x+k_z)/(4\pi)$, $z = (k_z+k_y)/(4\pi)$ changing from 0 to 1

To investigate temperature region of existence of a short-range order, the equations (6), (7) have been solved numerically for spin $S = 1/2$

(Figs. 3-5). In Fig. 3 the dependence of Neel temperature on $r$ in the mean-field theory, SWT, SSWT, and Tyablikov theory is shown. The mean-field theory completely neglects spin correlations and consequently does not describe frustration, as $T_N^{MF}$ is not strongly reduced near $r=1/2$. The Neel temperature in the other approaches is low, in particular in Tyablikov method $T_N^{Tyab}$ is lower, than in SSWT and SWT, except for a narrow vicinity of a frustration point. Though for non-frustrated antiferromagnets the result $T_N^{Tyab} \ll T_N^{SSWT}$ is physically meaningful and corresponds to better account of temperature fluctuations in Tyablikov theory in comparison with SSWT, near to a frustration point SSWT results are more reliable, since contrary to the Tyablikov theory, SSWT treats correctly quantum fluctuations [3].

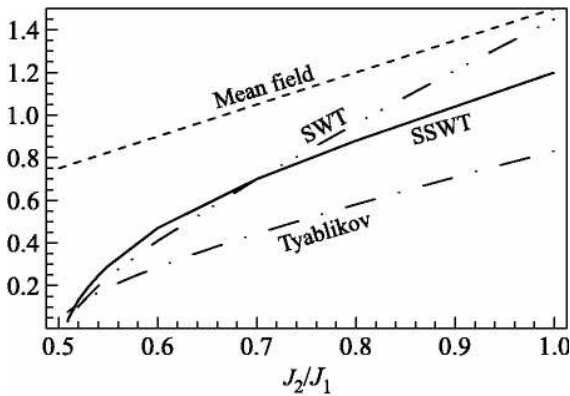

Fig. 3. Dependence of Neel temperature on $r = J_2/J_1$ in the mean-field theory, SWT, SSWT, and the Tyablikov theory.

On Fig. 4 the dependence of the sublattice magnetization (at $T<T_N$) and the chemical potential $\mu$ (at $T>T_N$) on temperature is shown for various $r$. Because of quantum fluctuations, which are especially strong near a frustration point $r = 1/2$, there is a strong suppression of the sublattice magnetization near this point. Temperature dependence of the chemical potential determining inverse correlation length, appears linear in this method. In Fig. 5 the behavior of average short-range order parameter for various values of $r$ is shown. One can see that at strong enough frustration there is a wide area of temperatures above $T_N$ with strong SRMO. For example, at $r = 0.53$ up to $T = 10T_N$ we find $\gamma > 1$ Thus, within SSWT it is shown, that there is a range of parametres of the model (1) where there a strong short-range order exist at $T \gg T_N$.

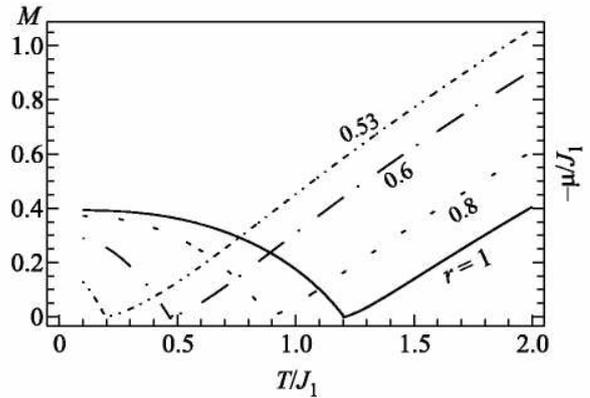

Fig. 4. Temperature dependence of the sublattice magnetisation ($T < T_N$) and chemical potential $\mu$ ($T > T_N$) for various $r$

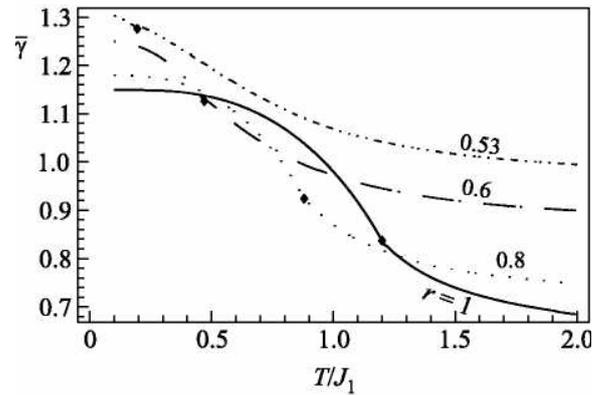

Fig. 5. Dependence of average SRMO parameter on temperature for various $r$. Points show the position of the Neel temperature

To apply the obtained results to $\gamma$-Fe, it is necessary to determine the model parametres $J_1$ and $J_2$. Methods of determination of the exchange integrals, based on band structure calculations, meet some difficulties which are connected, in particular, with non-Heisenberg exchange [10]. For example, exchange parametres strongly depend on the method of their evaluation. Nevertheless, band structure calculations give some arguments in favor of presence of strong magnetic frustration in $\gamma$-Fe [11]. We would like to note also, that degree frustration can be tuned by changing volume of an elementary cell (e.g. introducing impurities, etc, see [12]). Though experimental determination of the discussed parameters also meets a number of difficulties [13], it can confirm or deny the hypothesis of closeness of $\gamma$-Fe to a frustration point.

In summary, the examined approach can give the description $\alpha$-$\gamma$ transition in Fe, caused by magnetic fluctuations. Due to strong SRMO, transition from the FCC to BCC phase can take place at temperatures, substantially higher than $T_N$, which is in agreement with the experimental value of $T_C / T_N$ and can justify magnetic mechanism of structural transformation. Further extension of this approach (e.g. including potential of crystal lattice deformation) will allow describing quantitatively the discussed structural phase transition.

Authors express their gratitude to Yu.N.Gornostyriev for valuable discussions and comments on a problem and the content of the work. The work is support by grants of the Russian basic research foundation 07-02-01264a and 4640.2006.2 and the grant 49-07-01 (by "MMK", "Ausferr", and "Intels").

1. S. C. Abrahams, L. Guttman, and J. S. Kasper, Phys. Rev. **127,** 2052 (1962); G.J. Johanson, M.B. McGirr, and D.A. Wheeler, Phys. Rev. B 1, 3208 (1970); Y. Tsunoda, N. Kunitomi, and R. M. Nicklow, J. Phys. F: Met. Phys. **17,** 2447 (1987).
2. M. Takahashi, Phys. Rev. B **40**, 2494 (1989); D. J. Yoshioka, Phys. Soc. Jpn. **58**, 3733 (1989).
3. V. Yu. Irkhin, A. A. Katanin, and M. I. Katsnelson, Phys. Rev. B 60, 1082 (1999); A.A. Katanin and V. Yu. Irkhin, Sov. Phys. Uspekhii, **177,** 639 (2007).
4. V. Yu. Irkhin, A. A. Katanin, and M. I. Katsnelson, J. Phys.: Cond. Matter **4**, 5227 (1992).
5. O. P. Sushkov, J. Oitmaa, and Zheng Weihong, Phys. Rev. B 63, 104420 (2001); J. Sirker, Zheng Weihong, O. P. Sushkov, and J. Oitmaa, Phys. Rev. B **73**, 184420 (2006).
6. T. Yildirim, A. B. Harris, and E. F. Shender, Phys. Rev. B **58**, 3144 (1998).
7. V. G. Baryakhtar, V. N. Krivoruchko, and D.A. Yablonsky, Zh. Eksp.Teor. Fiz. **85**, 602 (1983); Green's Functions in the Theory of Magnetism [in Russian], Kiev, Naukova Dumka, 1984.
8. A. M. Tsvelik, *Quantum Field Theory in Condensed Matter Physics,* Cambridge University Press, Cambridge, 1998.
9. Th. Bruckel, B. Dorner, A. Gukasov, and V. Plakhty, Phys. Lett. A **162,** 357 (1992).
10. V. A. Gubanov, A. I. Liechtenstein, and A. V. Postnikov, *Magnetic and the Electronic Structure of Crystals,* Springer-Verlag, Berlin, Heidelberg, New York, 1992.
11. O. N. Mryasov et al., J. Phys.: Cond. Matter **3**, 7683 (1991); D. W. Boukhvalov, Yu. N. Gornostyrev, M. I. Katsnelson et al., Phys. Rev. Lett. **99**, 247205 (2007).
12. A. V. Ruban, M. I. Katsnelson, W. Olovsson et al., Phys. Rev. B 7**1,** 054402 (2005).
13. P. Boni, G. Shirane, and J. P. Wicksted, Phys. Rev. B **31**, 4597 (1985).